\documentclass[a4paper,11pt]{article}

\usepackage{jcappub}

\usepackage{a4}   
\usepackage{epsfig}
\usepackage{latexsym, amssymb} 
\usepackage{amsmath}
\usepackage{psfrag}
\usepackage{bm}
\usepackage{rotating}
\usepackage{hyperref}

\def\N{\mathcal{N}}

\def\U{{\bf U}} \def\n{\hat{n}} 
\def\t{\vec{\theta}} 
\def\k{{\bf k}}
\def\F{{\mathcal F}} 
\def\N{{\rm N}} \def\E{E} 
\def\snr1{ ${\rm SNR}_1$ } 
\def\P{{\mathcal P}}
\newcommand{\be}{\begin{equation}}
\newcommand{\e}{\end{equation}} \newcommand{\bear}{\begin{eqnarray}}
\newcommand{\ear}{\end{eqnarray}}

\title{Predictions for  BAO distance estimates from the
  cross-correlation of the  Lyman-$\alpha$ forest and redshifted
  21-cm emission.}

\author[a]{Tapomoy Guha Sarkar}
\author[b]{Somnath Bharadwaj}
\affiliation[a] {Department of Physics,  Birla Institute of Technology and Science, Pilani, Rajasthan, India.}
\affiliation[b]{Department of Physics and  Meteorology  \& Centre for
  Theoretical Studies, IIT, Kharagpur 721302, India.}
\emailAdd{tapomoy@pilani.bits-pilani.ac.in}
\emailAdd{somnath@phy.iitkgp.ernet.in}

\abstract{

We investigate the possibility of using the cross-correlation 
 of the Lyman-$\alpha$ forest and redshifted 21-cm emission
to detect the baryon acoustic oscillation (BAO). The standard Fisher
matrix formalism is used to determine the accuracy with which it will
be possible to measure cosmological distances using this signal.   
Earlier predictions \cite{dawson2012}  indicate   that it will be possible to 
measure the dilation factor $D_V$ with  $1.9 \%$ accuracy at $z=2.5$
from the BOSS Lyman-$\alpha$  forest auto-correlation. In this paper we 
investigate if it is  possible to improve the accuracy  using
the cross-correlation. 

We use a simple parametrization of the Lyman-$\alpha$ forest survey which
very loosely matches some properties  of the  BOSS.
For the redshifted 21-cm observations we consider a hypothetical radio interferometric array layout.
It is assumed that the observations span
$z=2$ to $3$  and covers the $10,000 \, {\rm deg}^2$  sky coverage of
BOSS. We find that  it is possible to
significantly increase the accuracy of the distance estimates  by
considering the cross-correlation signal.  
}
\begin{document}
\maketitle
\flushbottom

\section{Introduction}
Neutral hydrogen (HI) in the post-reionization epoch ($z < 6$) is
known to be an important cosmological probe seen in both emission and
absorption.  Here, the redshifted 21-cm emission \cite{hirev1,
  hirev3} and the transmitted QSO flux through the Lyman-$\alpha$
forest \cite{wien, Mandel} are both of utmost observational
interest. In a recent paper \cite{tgs5} have proposed the
cross-correlation of the 21-cm signal with the Lyman-$\alpha$ forest
as a new probe of the post-reionization era. While it is true that the
emission and the absorption signals both originate from neutral
hydrogen (HI) at the same redshift (or epoch), these two signals,
however, do not originate from the same set of astrophysical
sources. The 21-cm emission originates from the HI housed in the
Damped Lyman-$\alpha$ systems (DLAs) which are known to contain the
bulk of the HI at low redshifts \cite{proch05}. The collective
emission from the individual clouds appears as a diffuse background in
low frequency radio observations \cite{poreion2}. On the contrary,
the Lyman-$\alpha$ forest consists of a large number of Lyman-$\alpha$
absorption lines seen in the spectra of distant background
quasars. These absorption features arises due to small density
fluctuations in the predominantly ionized diffuse IGM. On large
scales, however, the fluctuations in the 21-cm signal and the
Lyman-$\alpha$ forest transmitted flux are both believed to be
excellent tracers of the underlying dark matter distribution. It has
been proposed  \cite{tgs5}   that the cross-correlation
between these two signals can be used to probe the power spectrum
during the post-reionization era. 

The HI power spectrum can be determined separately from observations
of the Lyman-$\alpha$ forest \cite{pspec} and the redshifted 21-cm
emission \cite{poreion1}.   The cross-correlation signal,
  however, has some unique features which makes it interesting to also
  consider this as a probe of the HI power spectrum in addition to the
  respective auto-correlation signals. In order to highlight this, we
  first briefly discuss some aspects of the individual
  auto-correlation signals.  First, it is only possible to detect the
  Lyman-$\alpha$ forest along discrete lines of sight to known
  quasars.  The Poisson noise arising from this discrete sampling is
  an important factor in limiting the accuracy to which the three
  dimensional auto-correlation power spectrum can be estimated from a
  given survey.  This limit depends on the mean quasar number density
  and the signal to noise ratios (SNR) of the individual quasar
  spectra which varies from quasar to quasar \cite{mcquinn2011}. Both
  of these are usually predetermined once the survey instrument and
  observational strategy are fixed, and it is typically not possible
  to improve these values further for a given survey.  In contrast,
  the redshifted 21-cm signal is sensitive to the HI distribution in
  the entire field of view.  The accuracy to which the
  auto-correlation power spectrum can be estimated is determined by
  the configuration of the radio-interferometric array and the system
  noise. It is, in principle, possible here to scale up the array and
  increase the observation time to reduce the system noise to a level
  where it is possible to measure the HI power spectrum at an accuracy
  that is comparable to the cosmic variance limit.  The redshifted
  21-cm signal, however, is buried in foregrounds from other
  astronomical sources like the galactic synchrotron emission,
  extra-galactic point sources, etc. Removing these foregrounds which
  are several orders of magnitude larger than the signal, poses a
  great challenge for detecting the post-reionization HI power
  spectrum \cite{fg10,fg11}.  The foregrounds in the redshifted 21-cm
  emission will not be correlated with the Lyman-$\alpha$ forest. We
  therefore expect the foreground problem to be much less severe for
  the cross-correlation signal. In fact, any residual foreground after
  subtraction will only contribute to the variance of the
  cross-correlation signal.  This is the key advantage of using the
  cross-correlation in comparison to the redshifted 21-cm
  auto-correlation. Comparing to the Lyman-$\alpha$ forest
  auto-correlation, the accuracy to which it is possible to estimate
  the cross-correlation power spectrum also is limited by the discrete
  QSO sampling. This dependence, however, is weaker for the
  cross-correlation as compared to the auto-correlation.  Given a
  Lyman-$\alpha$ forest survey, it may be possible to suitably design
  redshifted 21-cm observations such that the cross-correlation
  provides a more accurate estimate of the power spectrum in
  comparison to the Lyman-$\alpha$ forest auto-correlation. These two
  features, namely less severe foreground problems compared to the
  21-cm auto-correlation and the prospects of improving the SNR
  compared to the Lyman-$\alpha$ forest auto-correlation, provide
  motivation for considering the cross-correlation as a probe of the
  cosmological power spectrum.
 
Cosmological density perturbations drive acoustic waves in the
primordial baryon-photon plasma which are frozen once recombination
takes place at $ z \sim 1000$, leaving a distinct oscillatory signature
on the CMBR anisotropy power spectrum \cite{peeb70}. The sound horizon at
recombination  sets a standard ruler that maybe used to
calibrate cosmological distances.  Baryons contribute to 
 $15 \% $ of the total matter density,  and the baryon  acoustic
oscillations  are imprinted in the late time clustering of
non-relativistic matter. The signal, here, is however suppressed by a
factor $ \sim \Omega_b/\Omega_m \sim 0.1$,  unlike the  CMBR
temperature anisotropies where it is an order unity effect
\cite{komatsu}. The baryon acoustic oscillation (BAO) is  
a powerful  probe of cosmological parameters \cite{seoeisen,white}. 
This is particularly useful since the effect occurs on large scales
($\sim 150 \,  
\rm Mpc$), where the fluctuations are still in the linear regime. 
It is possible to measure the angular diameter distance $D_A(z)$ and 
the Hubble parameter $H(z)$  as functions of redshift using the  
the transverse and the longitudinal oscillations respectively. These 
provide   means for estimating cosmological
parameters and placing  stringent constraints on dark energy
models.  Several groups have  considered the   possibility of
detecting the BAO signal  using  redshift 21-cm emission
\cite{pen2008, maowu, masui10}.   
Recently  the BAO has been precisely measured at $z \sim 0.57$
\cite{2012MNRAS.427.3435A} using the galaxies in the SDSS~III  Baryon
Oscillation 
Spectroscopic Survey (BOSS; \cite{dawson2012}). 
The possibility of detecting the BAO signal in the Lyman-$\alpha$ forest 
has  been extensively studied  \cite{cosparam1}.  The BAO has
recently been detected at $z \sim 2.3$ \cite{busca2013,slosar2013}
using the  BOSS Lyman-$\alpha$ forest data. The combination $D_A^{0.2}
  \, H^{-0.8}$  has been  measured at a precision of $3.5 \%$. The
  data used in these works covers about one-third of the ultimate BOSS
  footprint. 

 The projected results for BOSS \cite{dawson2012} predict that at the
 end of the survey the overall dilation factor $D_A^{2/3} \, H^{-1/3}$
 shall be measured from the Lyman-$\alpha$ data at an accuracy level
 of $1.9 \%$ at $z \sim 2.5$. In this paper we investigate the
 possibility of improving the accuracy of the distance estimates by
 measuring the cross-correlation of the BOSS Lyman-$\alpha$ data with
 redshifted 21-cm maps. We propose a possible radio-interferometric
 array configuration and discuss the observational strategy required
 to achieve this.  The possibility of detecting the BAO using the
 cross-correlation signal has also been discussed in an earlier work
 \cite{2011arXiv1112.0745G} which investigates the possibility of
 detecting the BAO oscillatory feature in the cross-correlation
 multi-frequency angular power spectrum in the transverse and radial
 direction. The present work looks at the three dimensional cross
 power spectrum and makes predictions for the cosmological distance
 measures using a Fisher matrix analysis.

A brief  outline of the paper follows. 
 In Section 2  of this
paper we quantify the cross-correlation between the Lyman-$\alpha$ forest
and the 21-cm emission. In Section 3 we introduce an estimator 
for the cross-correlation signal and derive its statistical
properties. In Section 4 we consider the imprint of BAO on the cross
correlation signal and set up the Fisher matrix for theoretically
estimating the accuracy to which it will be possible to measure  
$D_A(z)$ and $H(z)$.  In section 5 we present  
several observational considerations and the strategy 
for measuring the
cross-correlation signal.  Finally, in  Section 6  we present
theoretical  prediction  
of  the accuracy at which it will be possible to measure 
 $D_A$ and  $H(z)$.  We have used a reference  cosmological model with cosmological
parameters $(\Omega_m h^2, \Omega_b h^2, \Omega_{\Lambda}, h, n_s,
\sigma_{8})= (0.136, 0.023, 0.726, 0.705, 
0.97, 0.82)$   \cite{komatsu} throughout this paper.

\section{The Cross-correlation Signal}
We quantify the fluctuations in the transmitted flux ${\F}(\n ,z)$
along a line of sight $\n$ to a quasar through the Lyman-$\alpha$
forest using $\delta_{\F}(\n ,z)= {\F}(\n ,z)/\bar{\F}-1$. For the
purpose of this paper we are interested in large scales where it is
reasonable to adopt the fluctuating Gunn-Peterson approximation
\cite{gunnpeter, bidav, pspec, pspec1}. This relates the transmitted
flux to the matter density contrast $\delta$ as
${\F}=\exp[-A(1+\delta)^{\kappa}]$  which does not include
  redshift space distortion, and therefore gives only a rough
  picture.  Here $A$ and $\kappa$ are two redshift dependent
quantities.  The function $A$ is of order unity \cite{bolton} and
depends on the mean flux level, IGM temperature, photo-ionization rate
and cosmological parameters \cite{pspec1}, while $\kappa$ depends on
the IGM temperature density relation \cite{mac, trc}. For our
analytic treatment of the Lyman-$\alpha$ signal, we assume that the
measured fluctuations $\delta_{\F}$ have been smoothed over a
sufficiently large length scale such that it is adequate to work with
only a linear term $\delta_{\F} \propto \, \delta$ \cite{bidav,
  pspec, pspec1, vielmat,saitta, slosar1}  which incorporates
  redshift space distortion and a possible bias (eq. \ref{eq:cc1}).
The terms of higher order in $\delta$ are expected to be important at
small length scales which have not been considered here.
 
We use $\delta_T(\n,z) $ to quantify the fluctuations in $T(\n,z) $
the brightness temperature of redshifted 21-cm radiation.  In the
redshift range of our interest ($z<3.5$), it is reasonable to assume
that $\delta_T(\n,z)$ traces $\delta$ with a possible bias
\cite{poreion1, poreion2}.  The bias is expected to be scale
dependent below the Jeans length-scale \cite{fang}. 
General non-linearity shall also make the bias scale dependent and
further, fluctuations in the ionizing background also give rise to a
scale dependent bias \cite{poreion3, poreion0}. This bias is
 moreover found to grow monotonically with $z$ \cite{marin}.
 However, numerical simulations 
\cite{bagla2, tgs2011}, indicate that it is adequate to use a constant, scale
independent bias at the large BAO scales of our interest.

 With the above mentioned assumptions and incorporating redshift space
 distortions we may express both $\delta_{\F}$ and $\delta_{T}$ as 
\be
 \delta_{\alpha}(\n,z) =  \int \ \frac{d^3\k}{(2\pi)^3} \ e^{i \k \cdot\n r}  
 C_{\alpha}[1 + \beta_{\alpha} \mu^2] \Delta(\k) \,.
\label{eq:cc1}
\e 
where $\alpha={\F}$ and $T$ refer to the Lyman-$\alpha$ forest and
21-cm signal respectively. Here $r$ is the comoving distance
corresponding to $z$, $\Delta({\bf{k}})$ is the matter density
contrast in Fourier space and $\mu= {\bf \hat{ k} \cdot \hat{n}}$.
 We adopt the  values $C_{\F}=-0.13$ and $\beta_{\F}= 1.58$ from the 
Lyman-$\alpha$ forest  simulations of \cite{mcd03} at $z=2.25$. 
\cite{McDonald2006}  have measured the 1-D power spectrum 
using the Lyman-$\alpha$ auto-correlation, and they present
estimates of the redshift  evolution of the bias in the  $z$ range $2$
to $3$.  This bias, however, has a significant contribution from the
statistical errors in the estimated mean flux  $\bar{\F}$ which
appears in the definition of $\delta_{\F}(\n ,z)$.  While this error 
causes a bias in the auto-correlation,  we do not expect it  to
 affect the cross-correlation considered here. To keep the analysis
 simple, we make the somewhat unrealistic assumption that bias for 
$\delta_{\F}(\n ,z)$  is constant across the redshift range of our
 interest. While including this is important in the real data
 analysis, we do not expect this to severely affect the predictions of
this paper.    We use  
$C_T=\bar{T} \, \bar{x}_{\rm HI} \, b$ and $\beta_T=f/b$ which can be
calculated for any $z$ for the 21-cm  signal  \cite{tgs5}. 
  We note that there are large uncertainties in the values of all the
  four parameters $C_T, C_{\F}, \beta_T$ and $ \beta_\F$ arising from
  our poor knowledge  of the state of the diffuse IGM and the
  systems that harbour bulk of the neutral hydrogen  at $z \sim 2.5$.

A QSO survey (eg. SDSS\footnote{http://www.sdss.org}), typically, covers a
large fraction of the entire sky.  In contrast, a radio
interferometric array (eg. GMRT\footnote{http://www.ncra.tifr.res.in} 
usually has a much 
smaller field of view ( $\sim 1^\circ$). Only the overlapping region
common to both these observations provides an estimate of the
cross-correlation signal.  We therefore use the limited field of view
$ L \times L $ ( $L$ in radians) of the radio telescope to estimate the
cross-correlation signal. Given this constraint,
it is a reasonable observational strategy to use several pointings of
the radio telescope to cover the entire region of the QSO survey. Each
pointing of the radio telescope provides an independent estimate of
the cross-correlation signal, which can be combined to reduce the
cosmic variance.

We have assumed that the field of view is sufficiently small ($L <<
1$) so that the curvature of the sky may be  ignored.  
In the flat sky
approximation,  the  unit vector $\n$ along  any line of sight can be
expressed as $\n = \hat{\bf{m}} + \vec{\theta} $, where $\hat{\bf{m}}$
is the line of sight to the centre of the field of view and $
\vec{\theta}$ is a two-dimensional ($2D$) vector on the plane of the
sky.   We further assume that the redshift range of the observation is  restricted to a 
relatively narrow  band $B$ centered at the redshift $z$.  We then have an 
$ L \ \times \ L \times \ B$  observational volume, and we use the
displacement  
$(\vec{\theta},\Delta z)$  relative to the center   as  observational
coordinates within 
 this volume. We have the  comoving displacement vector
 $\Delta {\bf r} = r \vec{\theta} + \Delta r \,  \hat{\bf{m}}$   
with $\Delta r = c \Delta z /H(z)$ 
corresponding to the observational coordinates $(\vec{\theta},\Delta z)$ .

It is convenient to decompose the observed
$\delta_{\F}(\vec{\theta},\Delta z)$ and 
 $\delta_T(\vec{\theta},\Delta z)$  into Fourier modes whereby 
\be \Delta_{\alpha}(\U, \tau ) = \int_{-B/2}^{B/2} \ d\, \Delta z \  \int
_{-L/2}^{L/2} \ d^2{\vec{\theta}} \  \ e ^{- 2 \pi i (\U \cdot
  \vec{\theta} + \tau \, \Delta z)} \ \delta_{\alpha}(\vec{\theta}, \Delta z )
\label{eq:ft}
\e 
where $\U$ is a two dimensional vector conjugate to
$\vec{\theta}$, and $\tau$ is conjugate to $\Delta z$. 

We use the power spectra  ${\P}_{\alpha \gamma}(U,\tau)$  to quantify the statistical 
properties of the observed fluctuation fields. These power spectra are 
defined through 
\begin{equation}
 {\rm Re}\langle 
\ \Delta_{\alpha}(\U, \tau) \Delta^{*}_{\gamma}(\U ', \tau')
\ \rangle = L^2 \, B \, \delta_{\U \U '}\  \delta_{\tau  \tau'}  \ \P_{\alpha \gamma} (U,\tau)
 \, , 
\label{eq:cc4}
\e 
and we have 
\begin{equation}
\P_{\alpha \gamma}(U,\tau)=F_{\alpha \gamma}(\mu) \, P(k)
\label{eq:cc5}
\e
where  $P(k)$ is the matter power spectrum and 
\be
F_{\alpha \gamma}(\mu)= \frac{H(z)}{r^2 c} C_{\alpha} C_{\gamma} \, (1 + 
\beta_{\alpha} \mu^2)  (1 + \beta_{\gamma} \mu^2)
\label{eq:cc6}
\e
Here $U$ refers to inverse angular scales and $\k_{\perp}=2 \pi \U/r$ refers to the component 
of the Fourier mode $\k$   perpendicular to the line of sight, while  $\tau$ refers to 
the  inverse  of $\Delta z$ and $k_{\parallel} = 2 \pi  H(z) \tau/c$ refers to the line of 
sight  
component of the  Fourier mode $\k$.  In the subsequent analysis we use 
 $\P_{\F T}(U,\tau)$ to  quantify the statistical properties of the 
cross-correlation signal.  
We had earlier, in Paper I \cite{tgs5}, used the multi-frequency  angular power 
spectrum (MAPS, \cite{datta1})  to quantify the cross-correlation 
signal. We note that the multi-frequency angular power spectrum $C_{\ell}(\Delta z)$
used earlier is the Fourier transform of the power spectrum   $\P_{\F T}(U,\tau)$
that we use here {\it ie.}
\begin{equation}
C_{\ell}(\Delta z)= \int d \tau \, e^{2 \pi i  \Delta z \, \tau} \,  \P_{\F T}(U,\tau)
\,.
\e
 with $\ell = 2 \pi U$. 

\section{The Cross-correlation Estimator}
 In this section we  construct an estimator for $P_{\F T}(U,\tau)$,   
and consider the statistical properties of this estimator. First, we
assume that both the Lyman-$\alpha$  forest and the 21-cm observations
are pixelized along  the $\Delta z$ axis into pixels or 
channels of width $\Delta z_c$ such that both $\delta_{\F}(\t,\Delta z)$ and 
$\delta_{T}(\t,\Delta z)$ are measured  only at discrete redshifts
$\Delta z_n= n \Delta z_c$ with $n=-N_c/2+1,
...,-2,-1,0,1,2,...,N_c/2$. Here $N_c+1$  is the  
total number of channels, and $(N_c+1) \, \Delta z_c=B$ is  
the total redshift interval or the bandwidth spanned by the
observations.  

 Considering first the Lyman-$\alpha$ forest, we have, till now,  considered
$\delta_{\F}(\t,\Delta z_n)$ as a continuous field defined at all points on
the sky. In reality, it is possible to measure this only  along a few,
discrete lines of sight where there are background quasars.  We 
account for  this by defining $ \delta_{{\F} o}(\vec{\theta},n) $. 
the observed fluctuation in the transmitted Lyman-$\alpha$ flux, as 
 \be \delta_{{\F} o}(\vec{\theta},n) =
\rho(\vec{\theta}) \, [ \, \delta_{\F}(\vec{\theta}, \Delta z_n) + \delta_{{\F}
  \N}(\vec{\theta},n) ] 
\label{eq:e1}
\e 
where $\delta_{{\F}  \N}(\vec{\theta}, n)$ is the contribution from
the pixel noise  in the quasar spectra and 
 \be \rho(\vec{\theta}) =\frac {\sum_a
  w_a \ \delta_D^{2}( \vec{\theta} - \vec{\theta}_a) }{\sum_a w_a} 
\label{eq:e2}
\e is the quasar sampling function. Here $a=1,2,...,N$ refers to the
different quasars in the $L \times L$ field of view, $\vec{\theta}_a$
and $w_a$ respectively refer to the angular positions and weights of
the individual quasars. We have the freedom of adjusting the weights
to suit our convenience.  It is possible to change the relative
contribution from the individual quasars by adjusting the weights
$w_a$.

The quasar sampling function $\rho(\vec{\theta})$ is zero everywhere
except  the angular  position of the different quasars.  It is
sometimes convenient to express the noise contribution in
eq. (\ref{eq:e1}) as
\be 
\rho(\vec{\theta}) \,  \delta_{{\F}
  \N}(\vec{\theta},n) = \frac{ 
\sum_a  w_a \ \delta_D^{2}( \vec{\theta} - \vec{\theta}_a) \
\delta_{{\F}  \N}(\vec{\theta}_a,n)}{\sum_a  w_a }
\label{eq:e3}
\e
where $\delta_{{\F}  \N}(\vec{\theta}_a,n)$ refers to the pixel noise
contribution for the different quasars. 
The faint quasars typically have
noisy spectra in comparison to the bright ones.  We can  take this
into account and choose the  weights $w_a$ so as to increase the
contribution from  the bright quasars relative to the faint ones,  
thereby maximizing  the SNR for  the signal estimator.   
For the present analysis we have
made the simplifying assumption that the magnitude of $\delta_{{\F}
  \N}(\vec{\theta}_a,n)$ is the same across all the quasars
irrespective of the quasar flux. We have modelled   $\delta_{{\F}
  \N}(\vec{\theta}_a,n)$ as Gaussian random variables with the noise
in the different pixels being uncorrelated {\it ie.}
\be \langle \delta_{{\F}  \N}(\vec{\theta}_a,n) \ \delta_{{\F}
  \N}(\vec{\theta}_b,m)  \rangle = \delta_{a,b} \delta_{n,m}
\sigma^2_{\F\N} 
\label{eq:e4}
\e
where $\sigma^2_{\F\N}$ is the variance of the pixel noise
contribution. It is appropriate to use uniform weights $w_a=1$ 
in this situation. 
Such an assumption is  justified in the  situation where  there exist 
high SNR measurements of the  transmitted  flux for all 
the quasars.

Considering next the sampling function $\rho(\vec{\theta})$, we assume
that the  quasars are randomly distributed with no correlation amongst
their angular position, and the positions  also being unrelated to
$\delta_{\F}$ and $\delta_{T}$.  In reality, the quasars do exhibit
clustering \cite{myers},  however the contribution from the Poisson
fluctuation is considerably more significant here and it is quite
justified to ignore the effect of quasar clustering for the present
purpose. The Fourier transform of $\rho(\vec{\theta})$ then has the
properties that 
\be 
\langle \tilde{\rho}(\U) \rangle = \delta _{\U,0}
\label{eq:e5}
\e
and 
\be 
\langle \tilde{\rho}(\U_1) \tilde{\rho}(\U_2)
\rangle = \frac{1}{N} \, \delta _{\U_1,\U_2} + \left( 1-\frac{1}{N} \right) 
\, \delta _{\U_1,0} \delta _{\U_2,0}
\label{eq:e6}
\
\e
which we shall use later. In the subsequent analysis, we also assume
that $N \gg1 $ whereby $(1-1/N) \approx 1$.

We use  $ {\Delta}_{{\F} o}(\U,\tau)$ to denote  the Fourier transform 
of $\delta_{{\F} o}(\vec{\theta}, \Delta z)$.  Using eq. (\ref{eq:e1}) we
have 
 \be 
{\Delta}_{{\F}  o}(\U,\tau) = \tilde{\rho}(\U) \otimes
[ {\Delta_{\F}}(\U, \tau) + \Delta_{\F \N}(\U, \tau)]
\label{eq:e7}
\e where  $ \otimes $ denotes a convolution defined as
\be 
\tilde{\rho}(\U) \otimes {\Delta_{\F}}(\U, \tau) =
\frac{1}{L^2} \sum_{\vec{U'}} \ \tilde{\rho}(\U - \vec{U'})
     {\Delta_{\F}}(\vec{U'}, \tau)
\label{eq:e8}
 \e 

Using eqs. (\ref{eq:cc5}), (\ref{eq:e3}),
(\ref{eq:e4}), (\ref{eq:e5}), (\ref{eq:e6})  and 
(\ref{eq:e7}) we calculate the following statistical properties of
$\Delta_{{\F} 0}$, 
\be 
 \langle \Delta_{{\F} o}(\U,\tau) \rangle =0
\label{eq:e9}
\e
and 
\begin{eqnarray} 
 \langle \Delta_{{\F} o}(\U_1,\tau_1)  \Delta^{*}_{{\F} o}(\U_2,\tau_2)  \rangle
&=&
\delta_{\U_1,\U_2}  \, \delta_{\tau_1,\tau_2}  \,\frac{B}{L^2}   \left[ \P_{\F \F}(\U_1, \tau_1) 
\right. \nonumber \\ &+& \left.  \frac{1}{n_Q} \left\{ L^{-2} \sum_{\U} 
\P_{\F \F}(\U,\tau_1)+ \Delta z_c \ \sigma^2_{\F \N }\right\}\right]
\label{eq:e10}
\end{eqnarray}
where  $\bar{n}_Q=N/L^2$  denotes the quasar density on the sky.

It is possible to simplify the sum over  $\U$ using Parseval's theorem
whereby 
\be 
\frac{1}{L^2}  \sum_{\U} \P_{\F \F}(\U, \tau) = p_{_{1\rm D}}(\tau)
\e 
and 
\be
p_{_{1\rm D}}(\tau)
=  \int \, d  \, \Delta z \ e^{2 \pi i \tau \, \Delta z} \xi_{\F}( \Delta z) \,.
\label{eq:e11}
\e
Here $\xi_{\F}(\Delta z) =\langle \delta_{\F}(\t_a,z_1) 
\delta_{\F}(\t_a,z_1+ \Delta z)  \rangle $ is the one dimensional (1D)
correlation function of the fluctuations in the transmitted flux along
 individual quasar spectra, and $p_{_1{\rm D}}(\tau)$ is the 1D power spectrum
corresponding to $\xi_{\F}(\Delta z)$.  The 1D correlation $\xi_{\F}(\Delta z)$, or
 equivalently $\xi_{\F}(v_{\parallel})$,  is traditionally used to
 quantify the Lyman-$\alpha$ forest along quasar spectra, and this has
 been quite extensively studied \cite{becker, coppolani, dodorico}.  

 We have 
\be 
 \langle \Delta_{{\F} o}(\U_1,\tau_1)  \Delta^{*}_{{\F} o}(\U_2,\tau_2)  \rangle
 = \delta_{\U_1,\U_2} \,  \delta_{\tau_1,\tau_2} \, L^{-2} \, B \, \P_{\F \F o}(\U_1, \tau_1 )
\label{eq:e12}
\e
where 
\be
\P_{\F \F o}(\U,\tau) =  \P_{\F \F}(U,\tau) 
+    \frac{1}{\bar{n}_Q} \left[ p_{_{1\rm D}}(\tau) + 
\Delta z_c \, \sigma^2_{\F \N} \right] 
\label{eq:e13}
\e 
 Here the term $\frac{1}{\bar{n}_Q} \left[ p_{_{1\rm D}}(\tau) +
  \Delta z_c \, \sigma^2_{\F \N} \right]$  arises due to the discrete
quasar sampling.  We  note that in our work the quasar density
$\bar{n}_Q$  has  been assumed to be  constant over the redshift range
of  interest.  In reality $n_Q$ will exhibit a redshift dependence
depending on the quasar  luminosity function and the magnitude limit
of the quasar survey \cite{jiang}, and it is necessary to take this
into account. 

Radio interferometric observations directly measure 
$\Delta_{To}(\U, \Delta z_n)$ which  are 
related to $\Delta_{To}(\U, \tau)$ through a Fourier transform
\begin{equation}
\Delta_{To}(\U, \tau)= \int \, d \, \Delta z \ e^{- 2 \pi i \Delta z \, \tau} 
\Delta_{To}(\U, \Delta z)  \,.
\label{eq:e14}
\e
We consider a
radio-interferometric array with  several antennas, each   of 
diameter $D$.  The antenna diameter and the field of view $L$
 are related   as $\lambda /D \approx L$, where
$\lambda$ is the observing wavelength. Each pair of antennas measures
$\Delta_T(\U,z_n)$  at a  particular $\U$ mode corresponding to
 $\U={\bf d}/\lambda$, where ${\bf d}$ is the antenna separation
 projected perpendicular to the line of sight. The 
baselines $\U$  corresponding to the different antenna pairs are, in
general,  arbitrarily distributed depending on the array
configuration.   The observed HI   fluctuation $\Delta_{To}(\U,n)$
at  two  different $\U$ values are correlated if $\mid \U_1-\U_2 \mid
\le  1/L$.  It is possible to combine the baselines where the signal
is correlated by binning  the $\U$ values  using cells of 
size $L^{-1} \times L^{-1}$. We then have the binned baselines at
$\U= (n_x \hat{i}+ n_y 
\hat{j} )/L$ ($n_x,n_y$ are integers)  which exactly match the
Fourier modes of the Lyman-$\alpha$ 
signal.  The HI signal at different $\U$ values are now uncorrelated.

We first considering  only $\Delta_{T {\N}}(\U, \Delta z_n)$  which is the noise
contribution to   $\Delta_{To}(\U, \Delta z_n)$. The 
noise in different channels and baselines is uncorrelated, and we have 
\be 
\langle \Delta_{{T\N } }(\U_1,\Delta z_n) \Delta^*_{{T} \N}(\U_2,\Delta z_m)
\rangle = \delta_{\U_1,\U_2} \delta_{n,m}\, L^2\, N_T(\U_1)
\label{eq:e15}
\e
Here 
$\N_T(\U)$ is the noise power spectrum. For a single polarization
and a single baseline, this is given by 
\be
\N_T(\U)=\left(\frac{T^2_{\rm sys}\,}{2 \,  \Delta
  \nu_c \,   \Delta t} \right) \ \frac{[ \int d \Omega \,
    {\mathcal P}(\t)]^2}{[\int d   \Omega \, {\mathcal P}^2(\t) ]}
\label{eq:e16}
\e 
where  $T_{\rm sys}$ is the system temperature, 
$\Delta \nu_c$ the frequency  interval corresponding to $\Delta z_c$,
$\Delta t$ the integration time and ${\mathcal P}(\t)$ is the
normalised  power  
pattern of the individual antennas \cite{gmrt}. 
The exact value of the ratio of the two integrals in
eq. (\ref{eq:e16}) depend on the antenna design. It is convenient here
to express  eq. (\ref{eq:e16})  as 
\be
\N_T(\U)=\frac{T^2_{\rm sys}\,  L^2}{\chi  \, N_{pol} \,M(\U) \,  \Delta
  \nu_c \,   \Delta t} \,.
\label{eq:e17}
\e 
where  $N_{pol}$ is the number of polarizations being used, $M(\U)$ the
number of baselines in the particular cell corresponding to $\U$,  and
$\chi$ is a factor whose value depends on the antenna beam pattern
${\mathcal P}(\t)$. For the purpose of this paper it is reasonable to
assume a value $\chi=0.5$. 

Writing 
 \be 
{\Delta}_{{T} o}(\U,\tau) = \Delta_{T}(\U,\tau) + \Delta_{T \N}(\U,\tau) 
\label{eq:e18}
\e 
we have 
\be 
\langle \Delta_{{T\N } }(\U_1,\tau_1) \Delta^*_{{T} \N}(\U_2,\tau_2)
\rangle = \delta_{\U_1,\U_2} \, \delta_{\tau_1,\tau_2} \, L^2\, B \,  \P_{T T o}(\U_1, \tau_1)
\label{eq:e19}
\e
where 
\be
\P_{T To}(\U,\tau)= \P_{TT}(U, \tau)  +   \Delta z_c \,
\N_T(\U)   \,.
\label{eq:e20}
\e

We finally consider the cross-correlation  for which we have 
 \be
 {\rm Re} \langle \frac{1}{2} \left[ {\Delta}_{{\F} o}(\U_1,\tau_1)
 {\Delta}^{*}_{{T} o}(\U_2,\tau_2) + 
   {\Delta}^{*}_{{\F} o}(\U_1,\tau_1) {\Delta}_{{T} o}(\U_2,\tau_2)
   \right] \rangle =  \delta_{\U_1,\U_2} \, \delta_{\tau_1,\tau_2}  B \, 
\P_{\F T o}(U_1,\tau_1 )
 \label{eq:e21}
\e
where 
\be 
\P_{\F T o}(\U,\tau) = \P_{\F T}(U, \tau)  \,.
 \label{eq:e22}
\e
 We use this to define  the estimator $ \hat{E}(\U,\tau)$ as 
\be
\E(U,\tau)= \frac{1}{2  B} 
\left[ \Delta_{{\F} o}(\U,\tau)  \Delta^{*}_{{T} o}(\U,\tau) +
   \Delta^{*}_{{\F} o}(\U,\tau)  \Delta_{{T} o}(\U,\tau)
   \right] 
 \label{eq:e23}
\e

The estimator has the property that  
\be
\langle \E(\U,p) \rangle =  \P_{\F T}(U, \tau) 
 \label{eq:e24} 
\e
{\it ie.} it is an unbiased estimator for the cross-correlation
signal. We next consider the covariance of the estimator
\be 
{\rm cov}[E(\U_1,\tau_1),E(\U_2,\tau_2)]=
\langle E(\U_1,\tau_1) \,
E(\U_2,\tau_2) \rangle - \langle E(\U_1,\tau_1) \rangle \, \langle
E(\U_2,\tau_2)\rangle 
 \label{eq:e25}
\e
which, we find, is diagonal 
\be 
{\rm cov}[E(\U_1,\tau_1),E(\U_2,\tau_2)]= \delta_{\U_1,\U_2} \, \delta_{\tau_1,\tau_2} 
\sigma^2[E(\U_1,\tau_1)]
 \label{eq:e26}
\e
and we need only consider the variance 
\be
\sigma^2[E(\U,\tau)] =
\langle E^2(\U,\tau) \rangle - \langle E(\U,\tau) \rangle^2 
 \label{eq:e27}
\e
which has a value 
\be
\Delta \P^2_{\F T}(\U,\tau) \equiv \sigma^2[E(\U,\tau)]=
\frac{1}{2} \left[ \P^2_{\F T}(U,\tau) +  \P_{\F \F o}(\U,\tau)  
 \P_{T T o}(\U,\tau) \right]
 \label{eq:e28}
\e

We use eq. (\ref{eq:e28}) along with eqs. (\ref{eq:e13}) and (\ref{eq:e20}) to calculate
the error in the estimated cross-correlation power spectrum.

\section{The Baryon Acoustic Oscillations}

The characteristic scale of the BAO is set by the sound horizon $s$ at
the epoch of recombination given by \be s = \int_a^{a_r}
\frac{c_s(a)}{a^2 H(a)} da \e where $a_r$ is the scale factor at the
epoch of recombination and $c_s$ is the sound speed given by $c_s(a) =
c/\sqrt{3(1 + 3 \rho_b/4 \rho_\gamma)}$, where $\rho_b$ and
$\rho_\gamma$ denotes the photon and baryonic densities respectively.
The comoving length-scale $s$ defines a transverse angular scale $
\theta_s = s [(1+z) D_A(z)]^{-1}$ and a radial redshift interval
$\Delta z_s =s H (z)/c$, where $D_A(z)$ and $H(z)$ are the angular
diameter distance and Hubble parameter respectively.  The comoving
length-scale $s= 143 \, \rm Mpc$ corresponds to $\theta_s=
1.38^{\circ}$ and $\Delta z_s = 0.07$ at $z = 2.5$.  Measurement of
$\theta_s$ and $\Delta z_s$ separately, allows the independent
determination of $ D_A(z)$ and $H(z)$.  Here we consider the determination of 
these two parameters from the BAO imprint on the cross-correlation signal. 
We now derive formulas to make error predictions for these parameters. 

We start with the Fisher matrix 
\be
F_{i j}= \sum_{\U} \sum_{\tau}
 \frac{1}{\Delta \P^2_{\F T}(\U,\tau)}
 \frac{\partial \P_{\F T}(\U,\tau)}{\partial q_i}
 \frac{\partial \P_{\F T}(\U,\tau)}{\partial q_j}
\label{eq:pe1}
\e
where $q_i$ refer to the cosmological parameters to be constrained.  

it is convenient to approximate the sums by integrals using
\be 
\sum_{\U} \sum_{\tau}  = \frac{1}{2} \int  \frac{d^2 \U}{L^{-2}} \,
 \int \frac{d \tau }{B^{-1}}=\frac{V}{2 (2 \pi)^3} \int \, d \k_{\perp} \, 
\int d k_{\parallel}
\label{eq:pe2}
\e
where $L^{-2}$ and $B^{-1}$ are the cell sizes for $\U$ and $\tau$ respectively, 
the factor $1/2$ is included to account for the fact that $(\U,\tau)$ and $(-\U,-\tau)$ 
do not  contain  independent information and $V=r^2 L^2 B c/H(z)$ is the observational 
volume. The integrals have limits defined by $\mid \U \mid \le U_{max}$  and 
$-(2 \Delta z_c)^{-1} \le \tau  \le (2 \Delta z_c)^{-1}$, or equivalently $\mid \k_{\perp} \mid 
\le 2 \pi U_{max}/r$ and $-c (2 \Delta z_c H(z))^{-1} \le k_{\parallel}
  \le c (2 \Delta z_c H(z))^{-1} $ where $U_{max}$ is the largest baseline in the 
radio interferometric array. We then have the Fisher matrix 
\be 
F_{i j}  = \frac{V}{(2 \pi)^3} \int \,  \, 
\frac{d^3  \k}{[\P^2_{\F T}(\k)+\P_{\F \F o}(\k) \P_{TTo}(\k)]}
 \frac{\partial \P_{\F T}(\k)}{\partial q_i}
 \frac{\partial \P_{\F T}(\k)}{\partial q_j}
\label{eq:pe3}
\e
which contains cosmological information from a variety of sources including
the redshift space distortion and the Alcock-Paczynski effect. In this 
work we would like to isolate the BAO constraints and ignore everything else. 
 This BAO information is 
mainly present  at small wave numbers  with the first peak at roughly
$ k \sim 0.045 {\rm Mpc}^{-1}$. The subsequent wiggles are 
 well suppressed by  $ k \sim 0.3 {\rm Mpc}^{-1}$ which is   within 
 the limits of the $\k_{\perp}$ and 
$k_{\parallel}$ integrals. It is thus quite justified to ignore the limits 
of the integral in eq. (\ref{eq:pe3}) and treat it as an integral over the 
entire $\k$ space.  Further, the BAO information is entirely contained in 
$P(k)$ and not $F_{\alpha \gamma}(\mu)$. We then have 
\be 
F_{i j}  = \frac{V}{(2 \pi)^2} \int \, k^2 \, dk  \, \int_{-1}^{1} \, d \mu \,  
\frac{1}{[P^2(k)+\P_{\F \F o}(\k) \P_{TTo}(\k)/F^2_{\F T}(\mu)]}
 \frac{\partial P(\k)}{\partial q_i}
 \frac{\partial P(\k)}{\partial q_j}
\label{eq:pe4}
\e
 
The subsequent analysis closely follows  \cite{seoeisen07}, and  we
refer the reader  
there for more details. We use $P_b=P-P_c$ to isolate the baryonic
features in the  
power spectrum, and we use this in the derivative $\partial
P(k)/\partial q_i$.   
Here $P_c$ refers to  the CDM power spectrum without any baryonic features. 
This gives 
\be 
P_b(\k) = \sqrt{8 \pi^2} A \, \frac{\sin x}{x} \, \exp \left[ -
  {\left(\frac{k}{k_{silk}} \right ) }^{1.4} \right] \, \exp \left [{ -
    \left( \frac{k^2}{2 k_{nl}^2}\right ) }\right] 
\e 
where ${k_{silk}}$ and ${k_{nl}}$ denotes the scale of `Silk-damping'  and 
`non-linearity' respectively.  In our analysis 
we have used $k_{nl} = (3.07 \, h^{-1} \rm Mpc)^{-1}$ and  $k_{silk} =
(7.76 \, h^{-1}\rm Mpc)^{-1}$  from \cite{seoeisen07}.  Here  
  $A$  is a normalization constant and 
$ x = \sqrt { k_{\perp}^2 s_{\perp}^2 + k_{\parallel}^2
  s_{\parallel}^2}$ where $ s_{\perp}$ and $ s_{\parallel}$
corresponds to $\theta_s$ and $\Delta z_s$ in units of distance. The 
value of  $s$ is known to a high level of precision from CMBR observations, 
and the values of $ s_{\perp}$ and $ s_{\parallel}$ are equal to $s$ for the reference values
of $D_A$ and $H(z)$. Changes in $D_A$ and $H(z)$ are reflected as changes in the 
values of $ s_{\perp}$ and $ s_{\parallel}$ respectively, and 
the fractional errors in $ s_{\perp}$ and $ s_{\parallel}$ correspond to fractional errors
in  $D_A$ and $H(z)$ respectively.  We use 
$q_1 = {\rm ln} (s_{\perp}^{-1})$ and $q_2 = {\rm ln}( s_{\parallel})$
as parameters in  
our analysis,  and determine the precision at which it will be
possible to constrain 
these using the location of the BAO features in the cross-correlation signal. 
Following \cite{seoeisen07} we have
the $2-D$ Fisher matrix 
 \be 
F_{ij} = V A^2 \int d k \int_{-1}^{1} d \mu 
\frac{k^2 \exp[-2 (k/k_{silk})^{1.4} - (k/k_{nl})^2]}
{[P^2(k)+\P_{\F \F o}(\k) \P_{TTo}(\k)/F^2_{\F T}(\mu)]} f_i(\mu) f_j(\mu) 
\label{eq:fmf}
 \e 
where  $f_1= \mu^2 -1$ and  $f_2= \mu^2$. We use  the Cramer-Rao bound 
$\delta q_i = \sqrt{F^{-1}_{ii}}$ to calculate the error in the parameter $q_i$.
 A combined distance measure $D_V$, also referred to as the
  ``dilation factor''  \cite{baoeisen05} 
\be
D_V(z)^3 = (1 + z) ^2 D_A(z) \frac{c z}{H(z)}
\e
is often used as a single  parameter to quantify BAO observations.
We use $\delta D_V/D_V = \frac{1}{3} ( 4 F^{-1}_{11} +
 4 F^{-1}_{12} +  F^{-1}_{22} )^{0.5}$ to calculate the relative 
 error in $D_V$. The dilation factor is particularly 
 useful when the sensitivity of the individual measurements of $D_A$
 and $H(z)$ is  low.

\section{Observational Considerations}
\label{sec:oc}
 The quasar redshift distribution peaks in the range  $z=2$ to $3$ 
\cite{sneider}, and   for our analysis we only consider the quasars 
in this redshift range.  For a quasar at a redshift $z_Q$, the region 
 $10,000 \, {\rm km \, s^{-1}}$ blue-wards of the quasar's Lyman-$\alpha$
emission  is excluded from the Lyman-$\alpha$ forest  due to the quasar's 
proximity effect.  Further, only the  pixels at least 1,000 $\rm km  \, 
s^{-1}$ red-ward of the quasar's Lyman-$\beta$ and O-$\rm VI$  lines 
are considered to avoid the possibility of confusing  the Lyman-$\alpha$ forest 
with the  Lyman-$\beta$ forest  or the intrinsic O-$\rm VI$ absorption.  For a 
quasar at the fiducial redshift $z_Q=2.5$, the 
above considerations would allow the Lyman-$\alpha$ forest to be measured
in the redshift range $ 1.96 \leq z \leq 2.39$ 
spanning an interval  $\Delta z =0.43$.  

We consider redshifted $21 \, {\rm cm}$  observations of bandwidth 
$ 128 \rm MHz$  covering  the frequency range
$ 355 \ \rm MHz$ to  $483 \rm MHz$ which corresponds to the 
redshift range $1.94\leq z \leq 3$. 
The Lyman-$\alpha$ forest of any particular quasar will be measured 
in a smaller interval $\Delta z \approx 0.4$ which is the deciding
factor for the cross-correlation signal. Thus, only a fraction 
(approximately $40 \%$) of the total number of quasars in this redshift
range $2$ to $3$ will contribute to the cross-correlation signal at 
any redshift $z$. We incorporate this in our  estimates by noting that 
$\bar{n}_Q$ in eq. (\ref{eq:e13}) refers to only $40 \%$ of all the quasars in 
the  entire $z$ range $2$ to $3$.

 We now discuss the kind of radio-interferometer that will be
  required.
The response of a radio-interferometric array falls at angular 
scales that are comparable to the field of view of the individual
antenna elements in the array, and it  is not sensitive to
angular scales that are larger than this field of view.   
The BAO scale subtends an angle of $1.56^{\circ}$ at $ z = 2.0$ and
smaller angles at higher redshifts.  In our analysis we consider a
radio-interferometric array which has antennas elements that are 
$2 \, {\rm m} \, \times \, 2 \, {\rm  m}$ in dimension. The array has
a  $L = 20^{\circ}$ field of view which  is considerably larger than
the BAO angular scale. 

We next discuss the array layout that would be required to observe 
the BAO features. In particular, we would like to determine the
antenna spacings that are required for the array to be sensitive
to the BAO features. The first  BAO peak is at roughly$ k_a \sim 0.045
{\rm Mpc}^{-1}$, followed by   subsequent wiggles whose amplitude
is  well suppressed within   $ k_b \sim 0.3 {\rm Mpc}^{-1}$.  
 Using $d/\lambda=U = k\, r/2\pi$, we estimate that the
Fourier modes $k_a$ and $k_b$ correspond to antenna separations $d_a =
31 $m and $d_b=207 $m respectively.  These figures roughly indicate the
range of antenna spacings  that would be required in the
radio-interferometric array.  Based on these considerations we
consider a radio interferometric array which has $N_{ann}$ antennas
distributed such that all the baselines $\vec{d}$ within $d_{max}=250
\, {\rm m}$ are uniformly sampled, whereby $M(\U)$ is independent of
$\U$ and we have $M(\U) \approx  (N_{ann}/250)^2$.  Using this in
eq. (\ref{eq:e17}), assuming $T_{sys}=100 \, {\rm K}$, $N_{pol}=2$,
$\chi=0.5$ we have 
\be N_{T}=1.0 \times 10^{-3} \, [m {\rm K}]^2 \,
\left( \frac{500}{N_{ann}} \right)^{2} \left( \frac{100 \, {\rm
    KHz}}{\Delta \nu_c} \right) \left( \frac{1000 \, {\rm hrs}}{\Delta
  t} \right) \,.
\label{eq:oc1}
\e We have $k_{max}=0.36 \, {\rm Mpc}^{-1}$ corresponding to
$d_{max}=250$m. The BAO features are well suppressed within $k_{max}$,
and this is large enough to capture the entire BAO information.   

The recent analysis of the BOSS Lyman-$\alpha$ forest auto-correlation
\cite{busca2013} has used pixels of width $210 \, {\rm km \,
  s^{-1}}$ for which they find a mean pixel SNR of $5.17$.
In our analysis we have assumed that the Lyman-$\alpha$ forest
and the redshifted 21-cm data are both   smoothed to this pixel size 
 which corresponds to $\Delta z_c=2.45 \times 10^{-3}$  and
 $\Delta \nu_c = 284 \, {\rm   kHz}$, or equivalently a radial
 separation of $2 \, h^{-1} \, {\rm Mpc}$   at $z=2.5$. We also assume
 that all the pixels have the same SNR of $5.17$.  Finally we note that 
the noise term in $\P_{\F \F o}$ also depends on $\xi_{\F}$ for which
we have used the values from \cite{becker}.

\section{Results}
 We  present the  results combining the
entire redshift range into a single bin centered at $z=2.5$. 
  We have considered a total sky coverage
  of $\sim 10,000 \rm deg^2$, which approximately matches the ultimate
  sky  
  coverage of BOSS and  which corresponds to 25 pointings of the
  $20^{\circ} \times 20^{\circ}$ field of view of the radio
  interferometer considered here.  The value of the BAO length scale
$s$ 
is well constrained from CMBR observations.  We have used the values
$s_\perp= s_\parallel = 1$ (in units of the BAO length scale $s = 143
\rm Mpc$) for the reference cosmological model.  We use the Fisher
matrix (eq. \ref{eq:fmf}) to predict the uncertainty in the parameters
$q_1 = \ln (s_{\perp}^{-1})$ and $q_2 = \ln( s_{\parallel})$, where
$\delta D_A/D_A =\mid \delta q_1 \mid$ and $\delta H/H =\mid \delta
q_2\mid $ . 

For the observational framework outlined in Section 5, the Fisher
matrix essentially depends on the two observational parameters
$\bar{n}_Q$ and $N_T$.  To recapitulate, $\bar{n}_Q$ refers to
approximately $0.4$ times the total angular density of quasars in the
redshift range $2$ to $3$, and $N_T$ is the system noise which depends
on the number of antennas $N_{ann}$ and the observing time $\Delta t$
(eq. \ref{eq:oc1}).  The limits $\bar{n}_Q \rightarrow \infty$ and
$N_T \rightarrow 0$, which correspond to $\P_{\F \F o} \rightarrow
\P_{\F \F}$ (eq. \ref{eq:e13}) and $\P_{T T o} \rightarrow \P_{T T}$
(eq. \ref{eq:e20}) respectively, set the upper bound for the SNR,
which also corresponds to the cosmic variance.
  In this limit,
  where the SNR depends only on the volume corresponding to the
 total field of view and the $B=1.0$ redshift
  interval,  we have $\delta
  D_V/D_{V} = 0.15 \%$,  $\delta H/H = 0.25 \%$  and   $D_A/D_{A} = 0.15
  \%$ which are independent of
  any   of the other  observational details.  Figure
\ref{fig:contour} shows contours of  $\delta D_V/D_{V}$ as a function
of  $\bar{n}_Q$ and $N_T$.  The fractional errors
  decrease very slowly  beyond $\bar{n}_{Q} > 50 \rm
  deg^{-2}$ or $N_T < 10^{-6} \rm mK^2$ at the bottom right corner of
  the panels, and  the  cosmic variance  limit is  reached
  beyond  the range shown in the figure. 

The parameter values $\bar{n}_{Q} \sim 6 \, \rm deg^{-2}$ and $N_T
\sim 4.7 \times 10^{-5} \ \rm mK^2$, we see, are adequate for a $1 \%$
accuracy, whereas $\bar{n}_{Q} \sim 2 \, \rm deg^{-2}$ and $N_T \sim
3 \times 10^{-3} \ \rm mK^2$ 
are adequate for a $\sim 10 \%$ accuracy in $D_V$ .  The $\delta
D_A/D_A$ and  $\delta H/H$  contours which are not shown here
exhibit a $\bar{n}_Q$ and $N_T$  
dependence which is very similar to $\delta D_V/D_V$.  Typically, it is  
possible to  measure $D_V$  at a higher level of precision compared to
$D_A$ and $H$ for the same set of observational parameters.

\begin{figure}[h]
\psfrag{NT}[c][c][1][0]{{\bf\large{ $N_{T}$}}}
\psfrag{n}[c][c][1][0]{{\bf\large{ $\bar{n}_Q \, (\rm deg^{-2})$}}}
\psfrag{a}[c][c][1][0]{{\bf\large{ $(\rm mK^2)$}}}
\psfrag{B}[c][c][1][0]{{\bf{$\frac{\delta D_{A}}{D_{A}}$}}}
\psfrag{H}[c][c][1][0]{{\bf{$\frac{\delta H}{H}$}}}
\psfrag{DV}[c][c][1][0]{{\bf{$\frac{\delta D_{V}}{D_{V}}$}}}

\begin{center}

\resizebox{270pt}{150pt}{\includegraphics{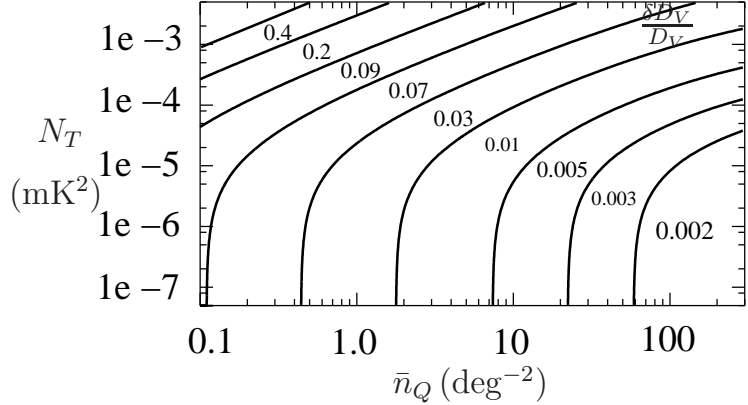}}

\end{center}
\caption{Contours of fractional errors in $D_V$
  at   $z = 2.5$.} 
\label{fig:contour}
\end{figure}

\subsection{Predictions for a BOSS-like survey}
\begin{figure}[h]
\psfrag{da}[c][c][1][0]{{\bf{${\delta D_{A}}/{D_{A}}$}}}
\psfrag{dv}[c][c][1][0]{{\bf{${\delta D_{V}}/{D_{V}}$}}}
\psfrag{dh}[c][c][1][0]{{\bf{${\delta H}/{H}$}}}
\psfrag{qq}[c][c][1][0]{{\bf {$\bar{n}_{Q} = 6.4 \rm deg^{-2}$}}}
\psfrag{mm}[c][c][1][0]{{\bf{$\bar{n}_Q = 1.0 $}}}
\psfrag{rr}[c][c][1][0]{{\bf{$\bar{n}_{Q} = 25.6 \rm deg^{-2}$}}}
\psfrag{pp}[c][c][1][0]{{\bf\Large{$\delta q_i$}}}
\psfrag{NT}[c][c][1][0]{{\bf\Large{$N_T$ ( $\rm mK^2$)}}}
\begin{center}
 \mbox{\epsfig{file=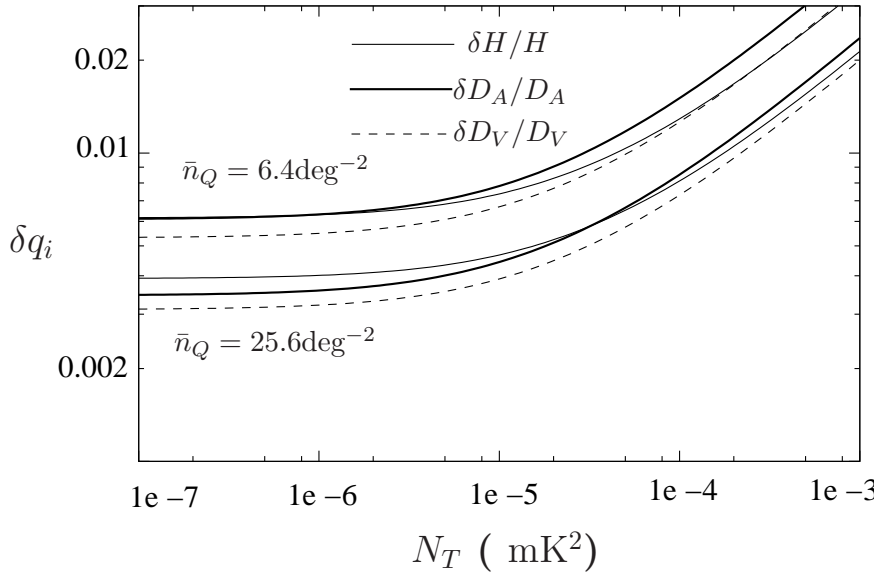, width=0.8550\textwidth,angle=0}}
\caption{The variation of the fractional error in the parameters for a
 as a function of $N_T$, for fixed
  $\bar{n}_Q$ corresponding to the BOSS and BIGBOSS like surveys. At large $N_T$ we
  have $\delta q_i \propto \sqrt{N_T}$ asymptotically as shown by the
  broken line.} 
\label{fig:varnt}
\end{center}
\end{figure}

The ongoing BOSS 
\cite{sneider,dawson2012} has a quasar number density of $\approx 16 \ \rm
deg^{-2}$ implying $\bar{n}_{Q} = 6.4 \ \rm deg^{-2}$ and is expected
to cover $\sim 10,000 \rm deg^2$ of the sky.  It has been predicted
that it will be possible to measure the dilation factor with an
accuracy of $\delta D_V/D_V = 1.9 \%$ at $z=2.5$ from the Lyman-$\alpha$
forest auto-correlation using BOSS \cite{dawson2012}.  We now
consider the cross-correlation signal for a BOSS-like survey. We
investigate if it will be possible to improve the accuracy of the
distance estimates by using the cross-correlation with redshifted
21-cm observations.

As outlined in Section~\ref{sec:oc}, we have used a simple
parametrization of the Lyman-$\alpha$ forest survey which very loosely
matches some properties of the ongoing BOSS.   
We first apply our simple model to calculate the
errors expected  in the distance estimates from the auto-correlation
signal for a BOSS-like survey. 
We use eq. (\ref{eq:fmf}) suitably modified for the Lyman-$\alpha$
forest auto-correlation. We find that our calculation predicts $\delta
D_V/D_V = 2.0 \%$ at $z=2.5$ which is in close agreement with the
detailed predictions \cite{dawson2012}. In the subsequent analysis we  
consider $\delta 
D_V/D_V =  2.0 \%$ as the fiducial error estimate for the
auto-correlation and investigate if it is possible to improve this
accuracy using the cross-correlation.

The sensitivity of the redshifted 21-cm observation is quantified by a
single parameter $N_T$.  Figure \ref{fig:varnt} shows how the relative
error in the distance estimates from the cross-correlation signal
varies 
with $N_T$.  We find that the errors scale as $\sqrt{N_T}$ for $N_T
\ge 10^{-4} {\rm mK}^2$.  It is possible to achieve the fiducial value 
$\delta D_V/D_V= 2.0 \%$ from the cross-correlation at $N_T = \, 2.9
\, \times \, 10^{-4}{\rm mK}^2$.  The error varies slower than
$\sqrt{N_T}$ in the range $N_T = 10^{-4}{\rm mK}^2$ to $N_T =
10^{-5}{\rm mK}^2$. We have $(\delta D_V/D_V,\delta D_A/D_A,\delta
H/H)=(1.3, 1.5, 1.3) \, \%$ and $(0.67, 0.78, 0.74) \, \%$ at $N_T =
10^{-4}{\rm mK}^2$ and at $N_T = 10^{-5}{\rm mK}^2$ respectively. The
errors do not significantly go down much further for $N_T <
10^{-5}{\rm mK}^2$, and we have $(0.55, 0.63, 0.63) \, \%$ at $N_T =
10^{-6}{\rm mK}^2$. We see that it is possible to significantly
increase the sensitivity relative to the BOSS Ly-$\alpha$
auto-correlation by considering the cross-correlation
 with redshifted 21-cm observations.    

\begin{figure}[h]
\psfrag{da}[c][c][1][0]{{\bf{$\frac{\delta D_{A}}{D_{A}}$}}}
\psfrag{blk}[c][c][1][0]{}
\psfrag{all}[c][c][1][0]{{\bf{$\frac{\delta H}{H}$}}}
\begin{center}
\resizebox{120pt}{120pt}{\includegraphics{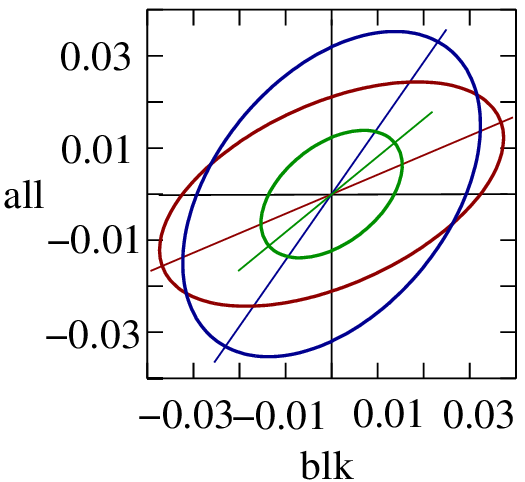}}
\resizebox{120pt}{120pt}{\includegraphics{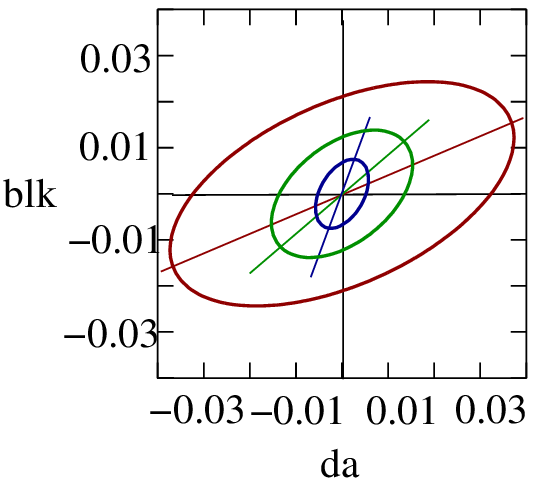}}
\resizebox{120pt}{120pt}{\includegraphics{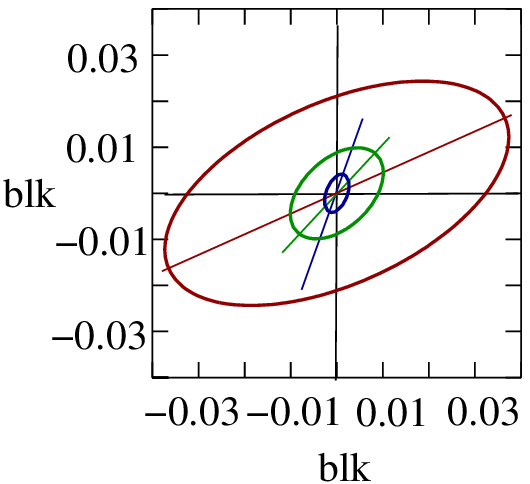}}
\end{center}
\caption{This shows $1-\sigma$ contours for the BOSS Lyman-$\alpha$
  forest auto-correlation (red), the redshifted 21-cm auto-correlation
  (blue) and the cross-correlation (green). The three panels  consider $N_T=
10^{-4} \,, 10^{-5} \, {\rm and} \,  10^{-6} \, {\rm mK}^2$ respectively
(left to right).  }
\label{fig:contourcomp}
\end{figure}

 Figure \ref{fig:contourcomp} shows a comparison of the relative
 errors in $D_A$ and $H$ for the three different estimates that can be
 obtained from  the BOSS Lyman-$\alpha$ forest data  and 21-cm
 observations, namely the  respective auto-correlations and the
 cross-correlation.  Note that each of  the three estimates  has a
 different sensitivity to the radial and transverse BAO signal,
 whereby  the  respective    $1-\sigma$ ellipses are  differently oriented.
The relatively large value of $\beta_{\F}$ for the  Lyman-$\alpha$
forest make it more sensitive to the radial clustering. In contrast,
the 21-cm  is more sensitive to the transverse signal
whereas  the cross-correlation  is oriented between the two
auto-correlations.  We find that it is possible to significantly
increase the sensitivity  relative to the BOSS   Lyman-$\alpha$ forest
auto-correlation by considering the  cross-correlation with 
21-cm observations. Note that it is particularly advantageous to use
the cross-correlation in a situation where the two auto-correlations
have comparable sensitivities ( $N_T = 10^{-4}{\rm mK}^2$ in this
case). In such a situation  we see that the cross-correlation provides
 more sensitive  distance estimates compared to both the individual
 auto-correlation. 

We now  discuss the configuration of the radio-interferometric array
and the observation  time that will be required.  It is necessary to
balance the number of antennas $N_{ann}$ and the total observing time
$T_{obs}$ which is divided over $25$ pointings {\it ie.} $\Delta t=
T_{obs}/25$ in eq. (\ref{eq:oc1}).  An array with $N_{ann}=2,000$
antennas will 
achieve  $N_T = 3.1 \times 10^{-5}{\rm  mK}^2$ with  $2$ years of
observation. With such an observation the cross-correlation  
will yield distance estimates with  $(\delta D_V/D_V,\delta
D_A/D_A,\delta H/H)=(1.2, 1.4, 1.3) \, \%$ which is a $1.7$ fold  
increase in the sensitivity relative to
the BOSS Lyman-$\alpha$ forest auto-correlation.   It is necessary to
increase 
the number of antennas for a significant improvement  beyond
this. Considering an array with $N_{ann}=4,000$ antennas, it will be
possible to achieve  $N_T = 3.9 \times 10^{-6}{\rm  mK}^2$ with  $4$
years of observation. This will yield distance estimates with
$(\delta D_V/D_V,\delta D_A/D_A,\delta H/H)=(0.64, 0.74, 0.71) \,
\%$ which is a $2.7$ fold  increase in the sensitivity relative to 
the BOSS Lyman-$\alpha$ forest auto-correlation.
Finally, we find that the sensitivity of the cross-correlation  
with the BOSS Lyman-$\alpha$ forest does not increase beyond 
$(\delta D_V/D_V,\delta D_A/D_A,\delta H/H)=(0.53,0.61,0.62) \,
\%$   irrespective of the number of antennas or the total observing
time.

The BIGBOSS \cite{bigboss} has been conceived as the successor to
BOSS.  BIGBOSS may achieve a quasar  density of $\sim 64 \, \rm
deg^{-2}$ which corresponds to $\bar{n}_Q=25.6 \ {\rm 
  deg}^{-2}$.  Our simple parametrization of the Lyman-$\alpha$ forest
predicts $\delta D_V/D_V= 0. 66 \, \%$ for the auto-correlation signal
assuming $5-\sigma$ SNR for all the spectra.   This implies  that the
sensitivity of the auto-correlation signal increase by a 
factor of $ \sim 3$ if the QSO number density is increased from  
$\bar{n}_{Q} = 6.4 \ \rm deg^{-2}$ to $\bar{n}_Q=25.6 \ {\rm 
  deg}^{-2}$. Considering the cross-correlation signal,  we 
have, for $N_T = 1.0 \times 10^{-5}{\rm  mK}^2$,   $(\delta D_V/D_V,\delta D_A/D_A,\delta H/H)=(0.39,0.44,0.47)\%$ . We find that there 
is a $\sim 1.7$ fold increase in  the  sensitivity of the cross-correlation signal 
if we consider BIGBOSS instead of BOSS. This increase in sensitivity is
approximately independent of the value of $N_T$. 

In conclusion, our calculation indicates that it is possible to
significantly increase the accuracy of the distance estimates  by
considering the cross-correlation signal.  However, more detailed and
realistic  modelling of the Lyman-$\alpha$ data and the 21-cm
observations are needed.   We expect the predictions of the  present 
 analysis to serve as a useful  signpost which indicates the 
direction for future work.

\section{Acknowledgement}
TGS would like to acknowledge Centre For Theoretical Studies, IIT
Kharagpur for using its various facilities. TGS would also like to
thank Tirthankar Roy Choudhury for useful discussions and help.
 
\bibliography{references}

\providecommand{\href}[2]{#2}\begingroup\raggedright\begin{thebibliography}{10}

\bibitem{dawson2012}
K.~S. {Dawson}, D.~J. {Schlegel}, C.~P. {Ahn}, S.~F. {Anderson},
  {\'E}.~{Aubourg}, S.~{Bailey}, R.~H. {Barkhouser}, J.~E. {Bautista},
  A.~{Beifiori}, A.~A. {Berlind}, V.~{Bhardwaj}, and D.~e.~a. {Bizyaev}, {\it
  {The Baryon Oscillation Spectroscopic Survey of SDSS-III}},  {\em \aj} {\bf
  145} (January, 2013) 10, [\href{http://xxx.lanl.gov/abs/1208.0022}{{\tt
  arXiv:1208.0022}}].

\bibitem{hirev1}
S.~R. {Furlanetto}, S.~P. {Oh}, and F.~H. {Briggs}, {\it {Cosmology at low
  frequencies: The 21 cm transition and the high-redshift Universe}},  {\em
  Physics Report} {\bf 433} (October, 2006) 181--301,
  [\href{http://xxx.lanl.gov/abs/astro-ph/}{{\tt astro-ph/}}].

\bibitem{hirev3}
M.~F. {Morales} and J.~S.~B. {Wyithe}, {\it {Reionization and Cosmology with 21
  cm Fluctuations}},  {\em ArXiv e-prints} (October, 2009)
  [\href{http://xxx.lanl.gov/abs/0910.3010}{{\tt arXiv:0910.3010}}].

\bibitem{wien}
D.~H. {Weinberg}, S.~{Burles}, R.~A.~C. {Croft}, R.~{Dave'}, G.~{Gomez},
  L.~{Hernquist}, N.~{Katz}, D.~{Kirkman}, S.~{Liu}, J.~{Miralda-Escude'},
  M.~{Pettini}, J.~{Phillips}, D.~{Tytler}, and J.~{Wright}, {\it {Cosmology
  with the Lyman-alpha Forest}},  {\em ArXiv Astrophysics e-prints} (October,
  1998) [\href{http://xxx.lanl.gov/abs/astro-ph/}{{\tt astro-ph/}}].

\bibitem{Mandel}
R.~{Mandelbaum}, P.~{McDonald}, U.~{Seljak}, and R.~{Cen}, {\it {Precision
  cosmology from the Lyman {$\alpha$} forest: power spectrum and bispectrum}},
  {\em \mnras} {\bf 344} (September, 2003) 776--788,
  [\href{http://xxx.lanl.gov/abs/astro-ph/}{{\tt astro-ph/}}].

\bibitem{tgs5}
T.~{Guha Sarkar}, S.~{Bharadwaj}, T.~R. {Choudhury}, and K.~K. {Datta}, {\it
  {Cross-correlation of the H I 21-cm signal and Ly{$\alpha$} forest: a probe
  of cosmology}},  {\em \mnras} {\bf 410} (January, 2011) 1130--1134,
  [\href{http://xxx.lanl.gov/abs/1002.1368}{{\tt arXiv:1002.1368}}].

\bibitem{proch05}
J.~X. {Prochaska}, S.~{Herbert-Fort}, and A.~M. {Wolfe}, {\it {The SDSS Damped
  Ly{$\alpha$} Survey: Data Release 3}},  {\em \apj} {\bf 635} (December, 2005)
  123--142, [\href{http://xxx.lanl.gov/abs/astro-ph/}{{\tt astro-ph/}}].

\bibitem{poreion2}
S.~{Bharadwaj}, B.~B. {Nath}, and S.~K. {Sethi}, {\it {Using HI to Probe Large
  Scale Structures at $z \sim 3$}},  {\em Journal of Astrophysics and
  Astronomy} {\bf 22} (March, 2001) 21--32,
  [\href{http://xxx.lanl.gov/abs/astro-ph/}{{\tt astro-ph/}}].

\bibitem{pspec}
R.~A.~C. {Croft}, D.~H. {Weinberg}, N.~{Katz}, and L.~{Hernquist}, {\it
  {Recovery of the Power Spectrum of Mass Fluctuations from Observations of the
  Ly alpha Forest}},  {\em \apj} {\bf 495} (March, 1998) 44--65,
  [\href{http://xxx.lanl.gov/abs/astro-ph/}{{\tt astro-ph/}}].

\bibitem{poreion1}
S.~{Bharadwaj} and S.~K. {Sethi}, {\it {HI Fluctuations at Large Redshifts: I
  Visibility correlation}},  {\em Journal of Astrophysics and Astronomy} {\bf
  22} (December, 2001) 293--307, [\href{http://xxx.lanl.gov/abs/astro-ph/}{{\tt
  astro-ph/}}].

\bibitem{mcquinn2011}
M.~{McQuinn} and M.~{White}, {\it {On estimating Ly{$\alpha$} forest
  correlations between multiple sightlines}},  {\em \mnras} {\bf 415} (August,
  2011) 2257--2269, [\href{http://xxx.lanl.gov/abs/1102.1752}{{\tt
  arXiv:1102.1752}}].

\bibitem{fg10}
A.~{Ghosh}, S.~{Bharadwaj}, S.~S. {Ali}, and J.~{Chengalur}, {\it {GMRT
  Observation Towards Detecting the Post-reionization 21-cm Signal}},  {\em
  Submitted to MNRAS} (2010).

\bibitem{fg11}
A.~{Ghosh}, S.~{Bharadwaj}, S.~{Saiyad Ali}, and J.~N. {Chengalur}, {\it
  {Improved foreground removal in GMRT 610 MHz observations towards redshifted
  21-cm tomography}},  {\em ArXiv e-prints} (August, 2011)
  [\href{http://xxx.lanl.gov/abs/1108.3707}{{\tt arXiv:1108.3707}}].

\bibitem{peeb70}
P.~J.~E. {Peebles} and J.~T. {Yu}, {\it {Primeval Adiabatic Perturbation in an
  Expanding Universe}},  {\em \apj} {\bf 162} (December, 1970) 815--825.

\bibitem{komatsu}
E.~{Komatsu}, J.~{Dunkley}, and M.~R. e.~a. {Nolta}, {\it {Five-Year Wilkinson
  Microwave Anisotropy Probe Observations: Cosmological Interpretation}},  {\em
  Astrophysical Journal Supplement} {\bf 180} (February, 2009) 330--376,
  [\href{http://xxx.lanl.gov/abs/0803.0547}{{\tt arXiv:0803.0547}}].

\bibitem{seoeisen}
H.~{Seo} and D.~J. {Eisenstein}, {\it {Probing Dark Energy with Baryonic
  Acoustic Oscillations from Future Large Galaxy Redshift Surveys}},  {\em
  \apj} {\bf 598} (December, 2003) 720--740,
  [\href{http://xxx.lanl.gov/abs/astro-ph/}{{\tt astro-ph/}}].

\bibitem{white}
M.~{White}, {\it {Baryon oscillations}},  {\em Astroparticle Physics} {\bf 24}
  (December, 2005) 334--344, [\href{http://xxx.lanl.gov/abs/astro-ph/}{{\tt
  astro-ph/}}].

\bibitem{pen2008}
T.~{Chang}, U.~{Pen}, J.~B. {Peterson}, and P.~{McDonald}, {\it {Baryon
  Acoustic Oscillation Intensity Mapping of Dark Energy}},  {\em Physical
  Review Letters} {\bf 100} (March, 2008) 091303,
  [\href{http://xxx.lanl.gov/abs/0709.3672}{{\tt arXiv:0709.3672}}].

\bibitem{maowu}
X.~{Mao} and X.~{Wu}, {\it {Signatures of the Baryon Acoustic Oscillations on
  21 cm Emission Background}},  {\em Astrophysical Journal Letters} {\bf 673}
  (February, 2008) L107--L110, [\href{http://xxx.lanl.gov/abs/0709.3871}{{\tt
  arXiv:0709.3871}}].

\bibitem{masui10}
K.~W. {Masui}, P.~{McDonald}, and U.~{Pen}, {\it {Near-term measurements with
  21 cm intensity mapping: Neutral hydrogen fraction and BAO at $z< 2 $}},
  {\em \prd} {\bf 81} (May, 2010) 103527,
  [\href{http://xxx.lanl.gov/abs/1001.4811}{{\tt arXiv:1001.4811}}].

\bibitem{2012MNRAS.427.3435A}
L.~{Anderson}, E.~{Aubourg}, S.~{Bailey}, D.~{Bizyaev}, M.~{Blanton}, A.~S.
  {Bolton}, J.~{Brinkmann}, J.~R. {Brownstein}, A.~{Burden}, A.~J. {Cuesta},
  L.~A.~N. {da Costa}, K.~S. {Dawson}, R.~{de Putter}, D.~J. {Eisenstein},
  J.~E. {Gunn}, H.~{Guo}, J.-C. {Hamilton}, P.~{Harding}, S.~{Ho},
  K.~{Honscheid}, E.~{Kazin}, D.~{Kirkby}, J.-P. {Kneib}, A.~{Labatie},
  C.~{Loomis}, R.~H. {Lupton}, E.~{Malanushenko}, V.~{Malanushenko},
  R.~{Mandelbaum}, M.~{Manera}, C.~{Maraston}, C.~K. {McBride}, K.~T. {Mehta},
  O.~{Mena}, F.~{Montesano}, D.~{Muna}, R.~C. {Nichol}, S.~E. {Nuza}, M.~D.
  {Olmstead}, D.~{Oravetz}, N.~{Padmanabhan}, N.~{Palanque-Delabrouille},
  K.~{Pan}, J.~{Parejko}, I.~{P{\^a}ris}, W.~J. {Percival}, P.~{Petitjean},
  F.~{Prada}, B.~{Reid}, N.~A. {Roe}, A.~J. {Ross}, N.~P. {Ross},
  L.~{Samushia}, A.~G. {S{\'a}nchez}, D.~J. {Schlegel}, D.~P. {Schneider},
  C.~G. {Sc{\'o}ccola}, H.-J. {Seo}, E.~S. {Sheldon}, A.~{Simmons}, R.~A.
  {Skibba}, M.~A. {Strauss}, M.~E.~C. {Swanson}, D.~{Thomas}, J.~L. {Tinker},
  R.~{Tojeiro}, M.~V. {Maga{\~n}a}, L.~{Verde}, C.~{Wagner}, D.~A. {Wake},
  B.~A. {Weaver}, D.~H. {Weinberg}, M.~{White}, X.~{Xu}, C.~{Y{\`e}che},
  I.~{Zehavi}, and G.-B. {Zhao}, {\it {The clustering of galaxies in the
  SDSS-III Baryon Oscillation Spectroscopic Survey: baryon acoustic
  oscillations in the Data Release 9 spectroscopic galaxy sample}},  {\em
  \mnras} {\bf 427} (December, 2012) 3435--3467,
  [\href{http://xxx.lanl.gov/abs/1203.6594}{{\tt arXiv:1203.6594}}].

\bibitem{cosparam1}
P.~{McDonald} and D.~J. {Eisenstein}, {\it {Dark energy and curvature from a
  future baryonic acoustic oscillation survey using the Lyman-{$\alpha$}
  forest}},  {\em \prd} {\bf 76} (September, 2007) 063009--063015,
  [\href{http://xxx.lanl.gov/abs/astro-ph/}{{\tt astro-ph/}}].

\bibitem{busca2013}
N.~G. {Busca}, T.~{Delubac}, J.~{Rich}, S.~{Bailey}, A.~{Font-Ribera},
  D.~{Kirkby}, J.-M. {Le Goff}, M.~M. {Pieri}, A.~{Slosar}, {\'E}.~{Aubourg},
  J.~E. {Bautista}, D.~{Bizyaev}, M.~{Blomqvist}, A.~S. {Bolton}, J.~{Bovy},
  H.~{Brewington}, A.~{Borde}, J.~{Brinkmann}, B.~{Carithers}, R.~A.~C.
  {Croft}, K.~S. {Dawson}, G.~{Ebelke}, D.~J. {Eisenstein}, J.-C. {Hamilton},
  S.~{Ho}, D.~W. {Hogg}, K.~{Honscheid}, K.-G. {Lee}, B.~{Lundgren},
  E.~{Malanushenko}, V.~{Malanushenko}, D.~{Margala}, C.~{Maraston},
  K.~{Mehta}, J.~{Miralda-Escud{\'e}}, A.~D. {Myers}, R.~C. {Nichol},
  P.~{Noterdaeme}, M.~D. {Olmstead}, D.~{Oravetz}, N.~{Palanque-Delabrouille},
  K.~{Pan}, I.~{P{\^a}ris}, W.~J. {Percival}, P.~{Petitjean}, N.~A. {Roe},
  E.~{Rollinde}, N.~P. {Ross}, G.~{Rossi}, D.~J. {Schlegel}, D.~P. {Schneider},
  A.~{Shelden}, E.~S. {Sheldon}, A.~{Simmons}, S.~{Snedden}, J.~L. {Tinker},
  M.~{Viel}, B.~A. {Weaver}, D.~H. {Weinberg}, M.~{White}, C.~{Y{\`e}che}, and
  D.~G. {York}, {\it {Baryon acoustic oscillations in the Ly{$\alpha$} forest
  of BOSS quasars}},  {\em Astronomy and Astrophysics} {\bf 552} (April, 2013)
  A96, [\href{http://xxx.lanl.gov/abs/1211.2616}{{\tt arXiv:1211.2616}}].

\bibitem{slosar2013}
A.~{Slosar}, V.~{Ir{\v s}i{\v c}}, D.~{Kirkby}, S.~{Bailey}, N.~G. {Busca},
  T.~{Delubac}, J.~{Rich}, {\'E}.~{Aubourg}, J.~E. {Bautista}, V.~{Bhardwaj},
  M.~{Blomqvist}, A.~S. {Bolton}, J.~{Bovy}, J.~{Brownstein}, B.~{Carithers},
  R.~A.~C. {Croft}, K.~S. {Dawson}, A.~{Font-Ribera}, J.-M. {Le Goff}, S.~{Ho},
  K.~{Honscheid}, K.-G. {Lee}, D.~{Margala}, P.~{McDonald}, B.~{Medolin},
  J.~{Miralda-Escud{\'e}}, A.~D. {Myers}, R.~C. {Nichol}, P.~{Noterdaeme},
  N.~{Palanque-Delabrouille}, I.~{P{\^a}ris}, P.~{Petitjean}, M.~M. {Pieri},
  Y.~{Pi{\v s}kur}, N.~A. {Roe}, N.~P. {Ross}, G.~{Rossi}, D.~J. {Schlegel},
  D.~P. {Schneider}, N.~{Suzuki}, E.~S. {Sheldon}, U.~{Seljak}, M.~{Viel},
  D.~H. {Weinberg}, and C.~{Y{\`e}che}, {\it {Measurement of baryon acoustic
  oscillations in the Lyman-{$\alpha$} forest fluctuations in BOSS data release
  9}},  {\em \jcap} {\bf 4} (April, 2013) 26,
  [\href{http://xxx.lanl.gov/abs/1301.3459}{{\tt arXiv:1301.3459}}].

\bibitem{2011arXiv1112.0745G}
T.~{Guha Sarkar} and S.~{Bharadwaj}, {\it {The Imprint of the Baryon Acoustic
  Oscillations (BAO) in the Cross-correlation of the Redshifted HI 21-cm Signal
  and the Ly-alpha Forest}},  {\em ArXiv e-prints} (December, 2011)
  [\href{http://xxx.lanl.gov/abs/1112.0745}{{\tt arXiv:1112.0745}}].

\bibitem{gunnpeter}
J.~E. {Gunn} and B.~A. {Peterson}, {\it {On the Density of Neutral Hydrogen in
  Intergalactic Space.}},  {\em \apj} {\bf 142} (November, 1965) 1633--1641.

\bibitem{bidav}
H.~{Bi} and A.~F. {Davidsen}, {\it {Evolution of Structure in the Intergalactic
  Medium and the Nature of the Ly alpha Forest}},  {\em \apj} {\bf 479} (April,
  1997) 523--532, [\href{http://xxx.lanl.gov/abs/astro-ph/}{{\tt astro-ph/}}].

\bibitem{pspec1}
R.~A.~C. {Croft}, D.~H. {Weinberg}, M.~{Pettini}, L.~{Hernquist}, and
  N.~{Katz}, {\it {The Power Spectrum of Mass Fluctuations Measured from the
  LYalpha Forest at Redshift Z=2.5}},  {\em \apj} {\bf 520} (July, 1999) 1--23,
  [\href{http://xxx.lanl.gov/abs/astro-ph/}{{\tt astro-ph/}}].

\bibitem{bolton}
T.~{Kim}, J.~S. {Bolton}, M.~{Viel}, M.~G. {Haehnelt}, and R.~F. {Carswell},
  {\it {An improved measurement of the flux distribution of the Ly{$\alpha$}
  forest in QSO absorption spectra: the effect of continuum fitting, metal
  contamination and noise properties}},  {\em \mnras} {\bf 382} (December,
  2007) 1657--1674, [\href{http://xxx.lanl.gov/abs/0711.1862}{{\tt
  arXiv:0711.1862}}].

\bibitem{mac}
P.~{McDonald}, J.~{Miralda-Escud{\'e}}, M.~{Rauch}, W.~L.~W. {Sargent}, T.~A.
  {Barlow}, and R.~{Cen}, {\it {A Measurement of the Temperature-Density
  Relation in the Intergalactic Medium Using a New Ly{$\alpha$} Absorption-Line
  Fitting Method}},  {\em \apj} {\bf 562} (November, 2001) 52--75,
  [\href{http://xxx.lanl.gov/abs/astro-ph/}{{\tt astro-ph/}}].

\bibitem{trc}
T.~R. {Choudhury}, T.~{Padmanabhan}, and R.~{Srianand}, {\it {Semi-analytic
  approach to understanding the distribution of neutral hydrogen in the
  Universe}},  {\em \mnras} {\bf 322} (April, 2001) 561--575,
  [\href{http://xxx.lanl.gov/abs/astro-ph/}{{\tt astro-ph/}}].

\bibitem{vielmat}
M.~{Viel}, S.~{Matarrese}, H.~J. {Mo}, M.~G. {Haehnelt}, and T.~{Theuns}, {\it
  {Probing the intergalactic medium with the Ly{$\alpha$} forest along multiple
  lines of sight to distant QSOs}},  {\em \mnras} {\bf 329} (February, 2002)
  848--862, [\href{http://xxx.lanl.gov/abs/astro-ph/}{{\tt astro-ph/}}].

\bibitem{saitta}
F.~{Saitta}, V.~{D'Odorico}, M.~{Bruscoli}, S.~{Cristiani}, P.~{Monaco}, and
  M.~{Viel}, {\it {Tracing the gas at redshift 1.7-3.5 with the Ly{$\alpha$}
  forest: the FLO approach}},  {\em \mnras} {\bf 385} (March, 2008) 519--530,
  [\href{http://xxx.lanl.gov/abs/0712.2452}{{\tt arXiv:0712.2452}}].

\bibitem{slosar1}
A.~{Slosar}, S.~{Ho}, M.~{White}, and T.~{Louis}, {\it {The acoustic peak in
  the Lyman alpha forest}},  {\em Journal of Cosmology and Astro-Particle
  Physics} {\bf 10} (October, 2009) 19--24,
  [\href{http://xxx.lanl.gov/abs/0906.2414}{{\tt arXiv:0906.2414}}].

\bibitem{fang}
L.~{Fang}, H.~{Bi}, S.~{Xiang}, and G.~{Boerner}, {\it {Linear evolution of
  cosmic baryonic medium on large scales}},  {\em \apj} {\bf 413} (August,
  1993) 477--485.

\bibitem{poreion3}
S.~{Wyithe} and A.~{Loeb}, {\it {Fluctuations in 21cm Emission After
  Reionization}},  {\em ArXiv e-prints} (August, 2007)
  [\href{http://xxx.lanl.gov/abs/0708.3392}{{\tt arXiv:0708.3392}}].

\bibitem{poreion0}
J.~S.~B. {Wyithe} and A.~{Loeb}, {\it {The 21-cm power spectrum after
  reionization}},  {\em \mnras} {\bf 397} (August, 2009) 1926--1934.

\bibitem{marin}
F.~{Marin}, N.~Y. {Gnedin}, H.~{Seo}, and A.~{Vallinotto}, {\it {Modeling The
  Large Scale Bias of Neutral Hydrogen}},  {\em ArXiv e-prints} (November,
  2009) [\href{http://xxx.lanl.gov/abs/0911.0041}{{\tt arXiv:0911.0041}}].

\bibitem{bagla2}
J.~S. {Bagla}, N.~{Khandai}, and K.~K. {Datta}, {\it {HI as a Probe of the
  Large Scale Structure in the Post-Reionization Universe}},  {\em ArXiv
  e-prints} (August, 2009) [\href{http://xxx.lanl.gov/abs/0908.3796}{{\tt
  arXiv:0908.3796}}].

\bibitem{tgs2011}
T.~{Guha Sarkar}, S.~{Mitra}, S.~{Majumdar}, and T.~R. {Choudhury}, {\it
  {Constraining large scale HI bias using redshifted 21-cm signal from the
  post-reionization epoch}},  {\em ArXiv e-prints} (September, 2011)
  [\href{http://xxx.lanl.gov/abs/1109.5552}{{\tt arXiv:1109.5552}}].

\bibitem{mcd03}
P.~{McDonald}, {\it {Toward a Measurement of the Cosmological Geometry at $z
  \sim 2$: Predicting Ly-{$\alpha$} Forest Correlation in Three Dimensions and
  the Potential of Future Data Sets}},  {\em \apj} {\bf 585} (March, 2003)
  34--51, [\href{http://xxx.lanl.gov/abs/astro-ph/}{{\tt astro-ph/}}].

\bibitem{McDonald2006}
P.~{McDonald}, U.~{Seljak}, S.~{Burles}, and D.~J. e.~a. {Schlegel}, {\it {The
  Ly{$\alpha$} Forest Power Spectrum from the Sloan Digital Sky Survey}},  {\em
  \apjs} {\bf 163} (March, 2006) 80--109,
  [\href{http://xxx.lanl.gov/abs/astro-ph/}{{\tt astro-ph/}}].

\bibitem{datta1}
K.~K. {Datta}, T.~R. {Choudhury}, and S.~{Bharadwaj}, {\it {The multifrequency
  angular power spectrum of the epoch of reionization 21-cm signal}},  {\em
  \mnras} {\bf 378} (June, 2007) 119--128,
  [\href{http://xxx.lanl.gov/abs/astro-ph/}{{\tt astro-ph/}}].

\bibitem{myers}
A.~D. {Myers}, R.~J. {Brunner}, G.~T. {Richards}, R.~C. {Nichol}, D.~P.
  {Schneider}, and N.~A. {Bahcall}, {\it {Clustering Analyses of 300,000
  Photometrically Classified Quasars. II. The Excess on Very Small Scales}},
  {\em \apj} {\bf 658} (March, 2007) 99--106,
  [\href{http://xxx.lanl.gov/abs/astro-ph/}{{\tt astro-ph/}}].

\bibitem{becker}
G.~D. {Becker}, W.~L.~W. {Sargent}, and M.~{Rauch}, {\it {Large-Scale
  Correlations in the Ly{$\alpha$} Forest at z = 3-4}},  {\em \apj} {\bf 613}
  (September, 2004) 61--76, [\href{http://xxx.lanl.gov/abs/astro-ph/}{{\tt
  astro-ph/}}].

\bibitem{coppolani}
F.~{Coppolani}, P.~{Petitjean}, F.~{Stoehr}, E.~{Rollinde}, C.~{Pichon},
  S.~{Colombi}, M.~G. {Haehnelt}, B.~{Carswell}, and R.~{Teyssier}, {\it
  {Transverse and longitudinal correlation functions in the intergalactic
  medium from 32 close pairs of high-redshift quasars}},  {\em \mnras} {\bf
  370} (August, 2006) 1804--1816,
  [\href{http://xxx.lanl.gov/abs/astro-ph/}{{\tt astro-ph/}}].

\bibitem{dodorico}
V.~{D'Odorico}, M.~{Viel}, F.~{Saitta}, S.~{Cristiani}, S.~{Bianchi},
  B.~{Boyle}, S.~{Lopez}, J.~{Maza}, and P.~{Outram}, {\it {Tomography of the
  intergalactic medium with Ly{$\alpha$} forests in close QSO pairs}},  {\em
  \mnras} {\bf 372} (November, 2006) 1333--1344,
  [\href{http://xxx.lanl.gov/abs/astro-ph/}{{\tt astro-ph/}}].

\bibitem{jiang}
L.~{Jiang}, X.~{Fan}, and R.~J.~e. {Cool}, {\it {A Spectroscopic Survey of
  Faint Quasars in the SDSS Deep Stripe. I. Preliminary Results from the
  Co-added Catalog}},  {\em Astrophysical Journal} {\bf 131} (June, 2006)
  2788--2800, [\href{http://xxx.lanl.gov/abs/astro-ph/}{{\tt astro-ph/}}].

\bibitem{gmrt}
J.~N. {Chengalur}, Y.~{Gupta}, and K.~S. {Dwarkanath}, {\em "Low Frequency
  Radio Astronomy"}.
\newblock NCRA-TIFR, India, 2007.

\bibitem{seoeisen07}
H.-J. {Seo} and D.~J. {Eisenstein}, {\it {Improved Forecasts for the Baryon
  Acoustic Oscillations and Cosmological Distance Scale}},  {\em \apj} {\bf
  665} (August, 2007) 14--24, [\href{http://xxx.lanl.gov/abs/astro-ph/}{{\tt
  astro-ph/}}].

\bibitem{baoeisen05}
D.~J. {Eisenstein}, I.~{Zehavi}, D.~W. {Hogg}, R.~{Scoccimarro}, M.~R.
  {Blanton}, R.~C. {Nichol}, R.~{Scranton}, and et~al., {\it {Detection of the
  Baryon Acoustic Peak in the Large-Scale Correlation Function of SDSS Luminous
  Red Galaxies}},  {\em \apj} {\bf 633} (November, 2005) 560--574,
  [\href{http://xxx.lanl.gov/abs/astro-ph/}{{\tt astro-ph/}}].

\bibitem{sneider}
D.~P. {Schneider}, P.~B. {Hall}, G.~T. {Richards}, and D.~E.~e. {Vanden Berk},
  {\it {The Sloan Digital Sky Survey Quasar Catalog. III. Third Data Release}},
   {\em Astrophysical Journal} {\bf 130} (August, 2005) 367--380,
  [\href{http://xxx.lanl.gov/abs/astro-ph/}{{\tt astro-ph/}}].

\bibitem{bigboss}
D.~J. {Schlegel}, C.~{Bebek}, H.~{Heetderks}, S.~{Ho}, and M.~{Lampton}, {\it
  {BigBOSS: The Ground-Based Stage IV Dark Energy Experiment}},  {\em ArXiv
  e-prints} (April, 2009) [\href{http://xxx.lanl.gov/abs/0904.0468}{{\tt
  arXiv:0904.0468}}].

\end{thebibliography}\endgroup
\end{document}